\documentclass[preprint,5p]{elsarticle}

\usepackage{hyperref} 
\usepackage{lineno}
\usepackage{color}
\modulolinenumbers[5]
\usepackage{todonotes}
\usepackage[separate-uncertainty=true]{siunitx}
\usepackage{amsmath}
\usepackage{booktabs}

\journal{Nuclear Instruments and Methods A}

\begin{document}

\begin{frontmatter}

\title{A beamline for fundamental neutron physics at TRIUMF }


\author[UM]{S. Ahmed}  
\author[UM]{T. Andalib}  
\author[CERN]{M.J.~Barnes}  
\author[UW,UM]{C.B.~Bidinosti}  
\author[TRIUMF]{Y.~Bylinsky} 
\author[TRIUMF]{J.~Chak} 
\author[UM]{M.~Das} 
\author[TRIUMF]{C.A.~Davis} 
\author[TRIUMF]{B. Franke} 
\author[UM]{M.T.W.~Gericke} 
\author[TRIUMF]{P. Giampa} 
\author[Coburg]{M.~Hahn}  
\author[UM]{S. Hansen-Romu} 
\author[RCNP]{K. Hatanaka} 
\author[UM,UW]{B.~Jamieson} 
\author[UBC]{D. Jones} 
\author[TRIUMF]{K. Katsika} 
\author[KEK]{S.~Kawasaki} 
\author[UM]{W. Klassen} 
\author[TRIUMF]{A. Konaka} 
\author[UNBC]{E.~Korkmaz} 
\author[TRIUMF]{F. Kuchler}  
\author[TRIUMF]{L.~Kurchaninov} 
\author[UM]{M. Lang} 
\author[TRIUMF,UM]{L. Lee} 
\author[TRIUMF,UW]{T.~Lindner} 
\author[UBC]{K.W.~Madison} 
\author[UM]{J.~Mammei} 
\author[TRIUMF,UW,UM]{R. Mammei} 
\author[UW,UM]{J.W.~ Martin} 
\author[TRIUMF]{R. Matsumiya} 
\author[TRIUMF,SFU]{R. Picker\corref{mycorrespondingauthor}}
\cortext[mycorrespondingauthor]{Corresponding author}  
\ead{rpicker@triumf.ca}
\author[TRIUMF,RCNP]{E.~Pierre} 
\author[TRIUMF]{W.D.~Ramsay} 
\author[TRIUMF]{Y.-N.~Rao} 
\author[TRIUMF]{W.R.~Rawnsley} 
\author[UM]{L. Rebenitsch} 
\author[TRIUMF]{C.A.~Remon} 
\author[TRIUMF]{W. Schreyer} 
\author[UW]{A. Sikora} 
\author[TRIUMF,SFU]{S.~Sidhu} 
\author[SFU]{J.~Sonier} 
\author[UW]{B. Thorsteinson} 
\author[UBC,TRIUMF]{S.~Vanbergen} 
\author[TRIUMF]{W.T.H.~van~Oers} 
\author[KEK]{Y. X. Watanabe} 
\author[TRIUMF]{D.~Yosifov} 

\address[UM]{University of Manitoba, Winnipeg, MB, Canada R3T 2N2}
\address[UBC]{The University of British Columbia, 2329 West Mall, Vancouver, BC, Canada V6T 1Z1}
\address[CERN]{Center for European Nuclear Research,CH-1211 Geneva 23, Switzerland}
\address[TRIUMF]{TRIUMF, 4004 Wesbrook Mall, Vancouver, BC V6T 2A3}
\address[UW]{University of Winnipeg, 515 Portage Avenue, Winnipeg, Manitoba R3B 2E9}
\address[Coburg]{University of Applied Science, Friedrich-Streib-Straße 2, 96450 Coburg, Germany}
\address[RCNP]{Research Center for Nuclear Physics, Osaka University, 10-1 Mihogaoka, Ibaraki, Osaka, 567-0047}
\address[Tokyo]{University of Tokyo, 7 Chome-3-1 Hongo, Bunkyō, Tokyo 113-8654, Japan}
\address[KEK]{High Energy Accelerator Research Organization, KEK,  1-1 Oho, Tsukuba, Japan, 305-0801}
\address[Kyoto]{Kyoto University, Yoshida-honmachi, Sakyo-ku, Kyoto 606-8501 Japan}
\address[Nagoya]{Nagoya University, Furo-cho, Chikusa-ku, Nagoya 464-8601, Japan}
\address[UNBC]{University of Northern British Columbia, 3333 University Way, Prince George, BC V2N 4Z9}
\address[SFU]{Simon Fraser University, 8888 University Drive, Burnaby, BC, Canada V5A 1S6}
 

\begin{abstract}
This article describes the new primary proton beamline 1U at TRIUMF.
The purpose of this beamline is to produce ultracold neutrons (UCN) for fundamental-physics experiments.
It delivers up to \SI{40}{\micro\ampere} of \SI{480}{\mega\electronvolt} protons from the TRIUMF cyclotron to a tungsten spallation target and uses a fast kicker to share the beam between the Center for Molecular and Materials Science and UCN.
The beamline has been successfully commissioned and operated with a beam current up to \SI{10}{\micro\ampere}, facilitating first large-scale UCN production in Canada.
\end{abstract}


\end{frontmatter}


\section{Introduction}

Ultracold neutrons (UCNs) have remarkably low kinetic energies ($< 300\,$neV). 
They can be stored in containers because they are reflected from suitable materials under all angles of incidence~\cite{zeldovich59}. 
This allows observation times of several hundred seconds and makes UCNs an ideal tool to study the neutron and its properties.
Examples are precise measurements of its decay lifetime, decay correlations, electric dipole moment, and interactions with gravity.

A large flux of free neutrons can be produced either in nuclear reactors~\cite{frm2, steyerl1986new,gatchina} or using spallation.
The latter usually uses proton beams of several hundred MeV impinging on high-Z targets~\cite{sinq, LANCSE2006,SNSspallation}.
Room-temperature moderators and reflectors like graphite, lead and heavy water are then used to maximize the thermal neutron flux inside a cold-neutron moderator material such as solid heavy water at \SI{10}{\kelvin}~\cite{rcnp} or liquid deuterium at \SI{20}{\kelvin}~\cite{snscoldneutron,steyerl1986new}.
To enhance the fraction of ultracold neutrons above the Maxwellian spectrum, the cold neutrons can either be Doppler shifted to lower energies, a process observing the restrictions of Liouville's theorem~\cite{steyerl1986new};
or they are subjected to specials materials, such as solid deuterium~\cite{SAUNDERS2004} or liquid helium~\cite{GOLUB1977}, where cold neutrons can excite collective excitations, lose almost all their kinetic energy and become ultracold.
The reverse process can be suppressed by cooling these so called superthermal converters to low temperatures.
At TRIUMF, superfluid $^4$He at around $1\,$K fulfills this role through phonon
and roton transitions.

A new proton beamline has been built at TRIUMF over the last several years to service the new UCN source built by the TUCAN (\textbf{T}RIUMF \textbf{U}ltraCold \textbf{A}dvanced \textbf{N}eutron) collaboration.
This paper will describe the design, construction and commissioning of this beamline.
We will also briefly introduce the UCN source and first results.

\section{Beamline 1 at TRIUMF}
\begin{figure*}[htb]
	\centering
	 \includegraphics[height=\textwidth, angle=90, trim = {0.5cm 0cm 0cm 0cm}, clip]{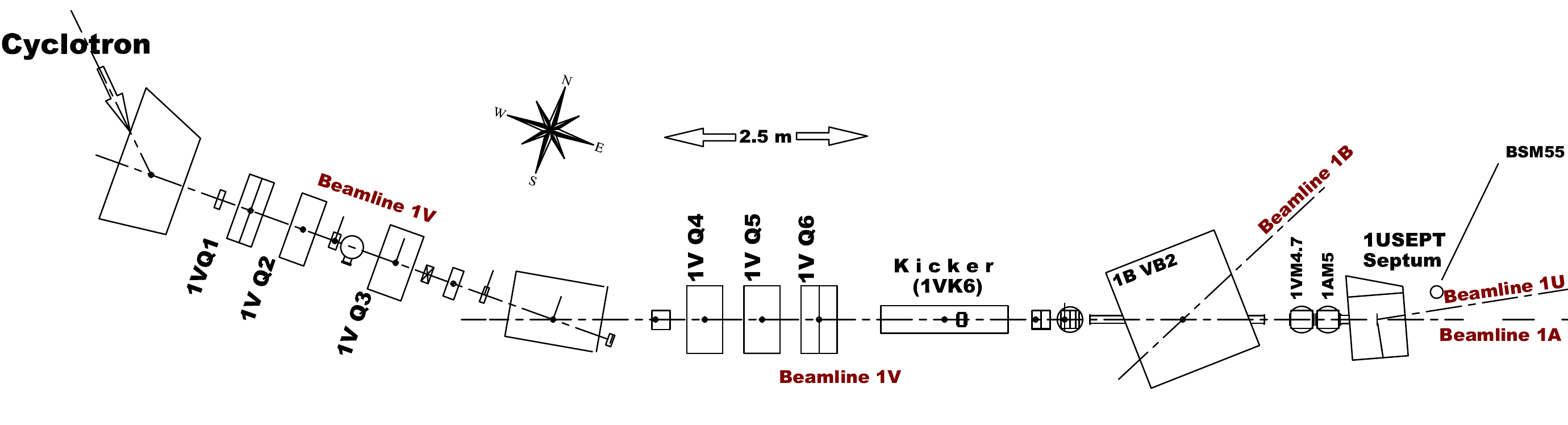}
	\caption{Schematic of beamline 1 at TRIUMF with beamlines 1B and 1U branching off at 1BVB2 and 1USEPT, respectively.}
	\label{fig:BL1Ulayout}
\end{figure*}
The TRIUMF cyclotron accelerates $H^-$ ions.
To extract particles, both electrons are stripped off using carbon foils mounted inside the cyclotron.
For beamline 1 (BL1) stripping happens at a radius of 
7.8
\,m 
from the center of the cyclotron, corresponding to a kinetic energy of \SI{480}{\mega \electronvolt}.
A proton current of up to \SI{120}{\micro \ampere} can be injected into BL1 in this way.
Before the installation of beamline 1U (BL1U), two operation modes of the beamline were possible depending on the setting of the dipole bender 1B VB2, see Fig.~\ref{fig:BL1Ulayout}.
At nominal magnet excitation, protons are diverted by  $43^{\circ}$ into beamline 1B for proton irradiation studies~\cite{blackmore2000operation}.
If 1B VB2 is turned off, the protons are injected into beamline 1A which serves two pion-production targets (1A T1, see Fig.~\ref{fig:BL1Ulayoutdown} and T2, made of beryllium) and a neutron-irradiation facility that doubles as beam dump~\cite{blackmore2014intensity}.
T1 and T2 feed the Center for Molecular and Material Science (CMMS) with several secondary beamlines providing experiments with mainly surface muons from pion decay~\cite{CMMS}.
To produce spallation neutrons, a new beamline called 1U, was installed between 2013 and 2016. 
The \SI{120}{\micro\ampere} proton beam can be shared between BL1A and BL1U by means of a fast kicker magnet.

Beamline 1U is \SI{16}{\meter} long, given by the space available in the Meson hall at TRIUMF.
Two additional constraints limit the space for beamline elements along its length.

One section of BL1U is located underneath the TRIUMF M15 channel~\cite{M15}, that extracts muons from the target 1AT1 in beamline 1A.
During construction of the beamline, a roughly \SI{4}{\meter}-long hole had to be cored into a shielding block pinned down by M15 to accommodate the 1U beam pipe, see Fig.~\ref{fig:BL1Ulayoutdown}, with no space for diagnostic elements.

The last section of the beamline is enclosed by a split shielding block with cutouts for the beamline and target arm (scallop block) to reduce the radiation from the spallation target to the upstream beamline part (see also Fig.~\ref{fig:target}).
As laid out in the following section, some beamline elements had to be designed to fit into a beam tube inside that shielding block.

\section{Beamline 1U Hardware}
\begin{figure*}[htb]
	\centering
	 \includegraphics[height=0.8\textwidth, angle=90, trim = {0cm 0cm 0cm 0cm}, clip]{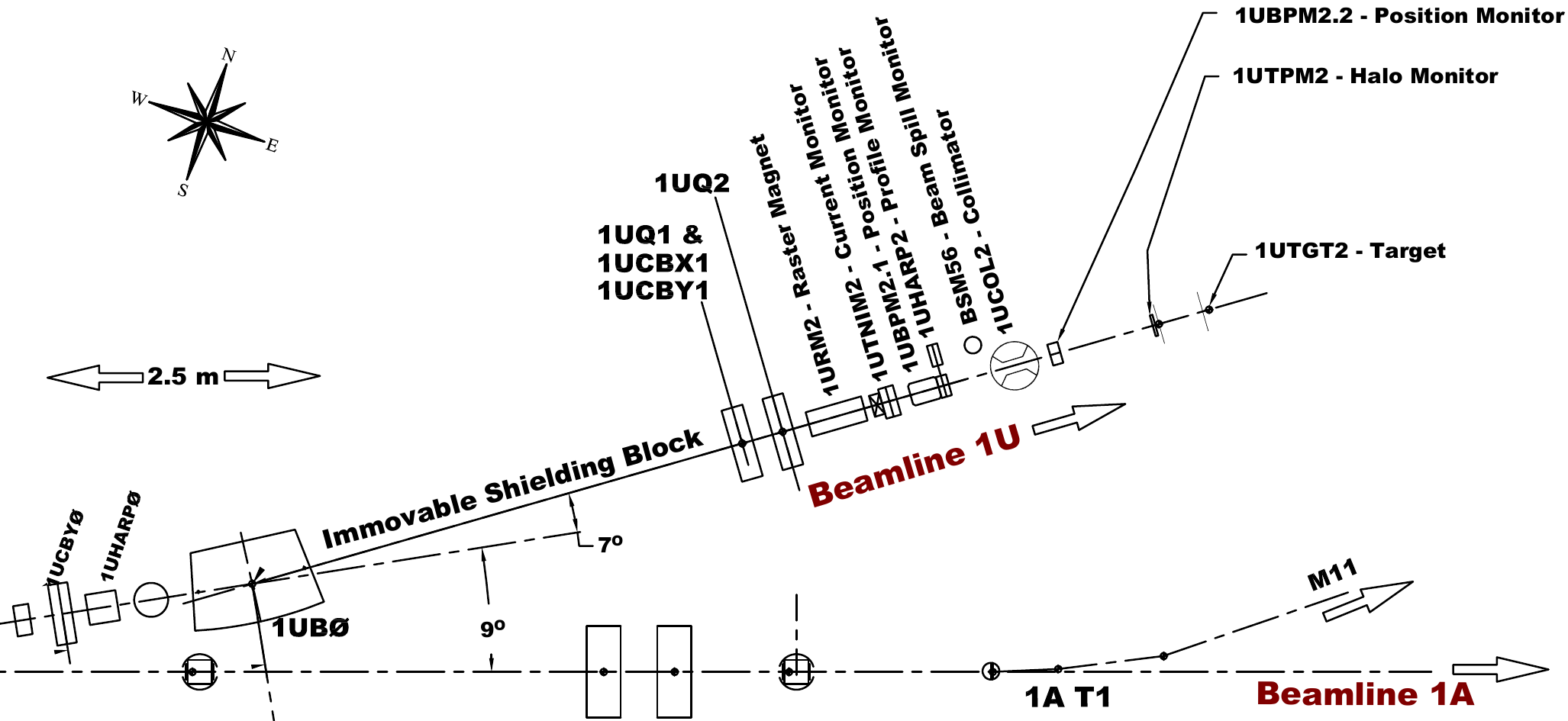}
	\caption{Schematic of BL1U at TRIUMF downstream of the septum 1USEPT.}
	\label{fig:BL1Ulayoutdown}
\end{figure*}
This section describes the main elements of beamline 1U as shown in Figs.~\ref{fig:BL1Ulayout} and \ref{fig:BL1Ulayoutdown}.
Tab.~\ref{tab:elements} summarizes the beamline elements mentioned in the text.
\begin{table}
\caption{Beamline 1U elements in alphabetical order.}
\begin{tabular}{lll}
\hline
Beamline element & Type  \\
\hline
1AM4.7 & beam position monitor \\
1AM5 & HARP wire chamber \\
1BVB2 & dipole bender \\
1UB0 & bending dipole \\
1UBPM2.1, 1UBPM2.2 & beam position monitors \\
1USEPT & septum magnet \\
1UCBY0 & vertical correction steerer \\
1UCBX1, 1UCBY1 & horiz. and vert. steerers \\
1UCOL2 & collimator \\
1UHARP0, 1UHARP2 & HARP wire chamber \\
1UQ1-Q2 & quadrupole doublet \\
1URM2 & raster magnet (planned) \\
1UTNIM2 & current monitor \\
1UTNPM2 & beam halo monitor \\
1VQ1-Q6 & two quadrupole triplets  \\
1VK6 & kicker magnet  \\
1VM4 & notch monitor \\
BSM55, BSM56 & beam spill monitor \\
\hline
\end{tabular}
\label{tab:elements}
\end{table}

\subsection{Fast kicker magnet for beam sharing}
\begin{figure}[htb]
	\centering
	\includegraphics[width=\columnwidth, trim = {0cm 0cm 6cm 0cm}, clip]{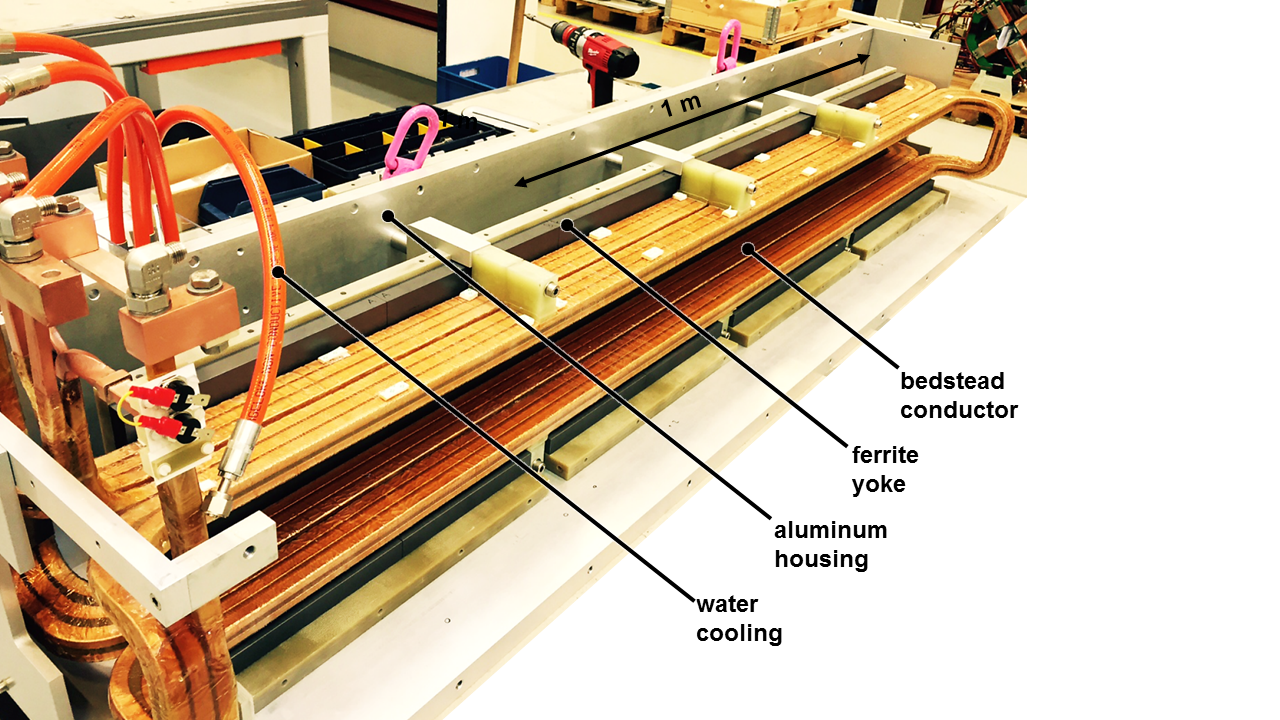}  	 
	\caption{Assembly of the kicker magnet at Danfysik.}
	\label{fig:kickerdanfysik}
\end{figure}
The $H^-$ injection system of the cyclotron utilizes a pulser with a variable duty cycle.
The repetition rate is $1.126\,$kHz and the pulse duration can be varied continuously between zero and \SI{878}{\micro\second} with a \SI{10}{\micro\second} minimum interruption between pulses.
This beam-off period is used to switch the beam between beamlines 1A and 1U.
The kicker magnet 1VK6 ramps up from zero to nominal current (\SI{193}{\ampere}) in \SI{50}{\micro\second}  creating a field integral of $0.0436\,$T$\cdot$m and deflecting protons upwards by $12\,$mrad ($0.68^{\circ}$).
The duty cycle has to be less than 94\% not to spill beam on the yoke of the septum magnet 1USEPT downstream.
The target of beamline 1U is designed to withstand a proton current of up to \SI{40}{\micro\ampere} which corresponds to one 
\SI{120}{\micro\ampere} proton pulse deflected into beamline 1U and two pulses transmitted undeflected to beamline 1A (a kick fraction of $1/3$).
Therefore, the kicker magnet and power supply has a maximum repetition rate of \SI{400}{\hertz}.
Any combination of deflected and undeflected pulses can be produced for other sharing ratios between the two beamlines.

The kicker magnet and power supply have been built by Danfysik~\cite{danfysik}.
Inspired by the MedAustron configuration~\cite{barnes}, the magnet as shown in Fig.
~\ref{fig:kickerdanfysik} consists of so-called bedstead conductors made of copper, a ferrite yoke and an aluminum box~\cite{Hahn2012}.
Steel end plates reduce the fringe magnetic field.
The kicker power supply provides up to \SI{1.8}{\kilo \volt} to energize (or de-energize) the magnet within \SI{50}{\micro\second}.

\subsection{Capacitive Probe}
\begin{figure}
	\centering
    \includegraphics[width=0.9\columnwidth]{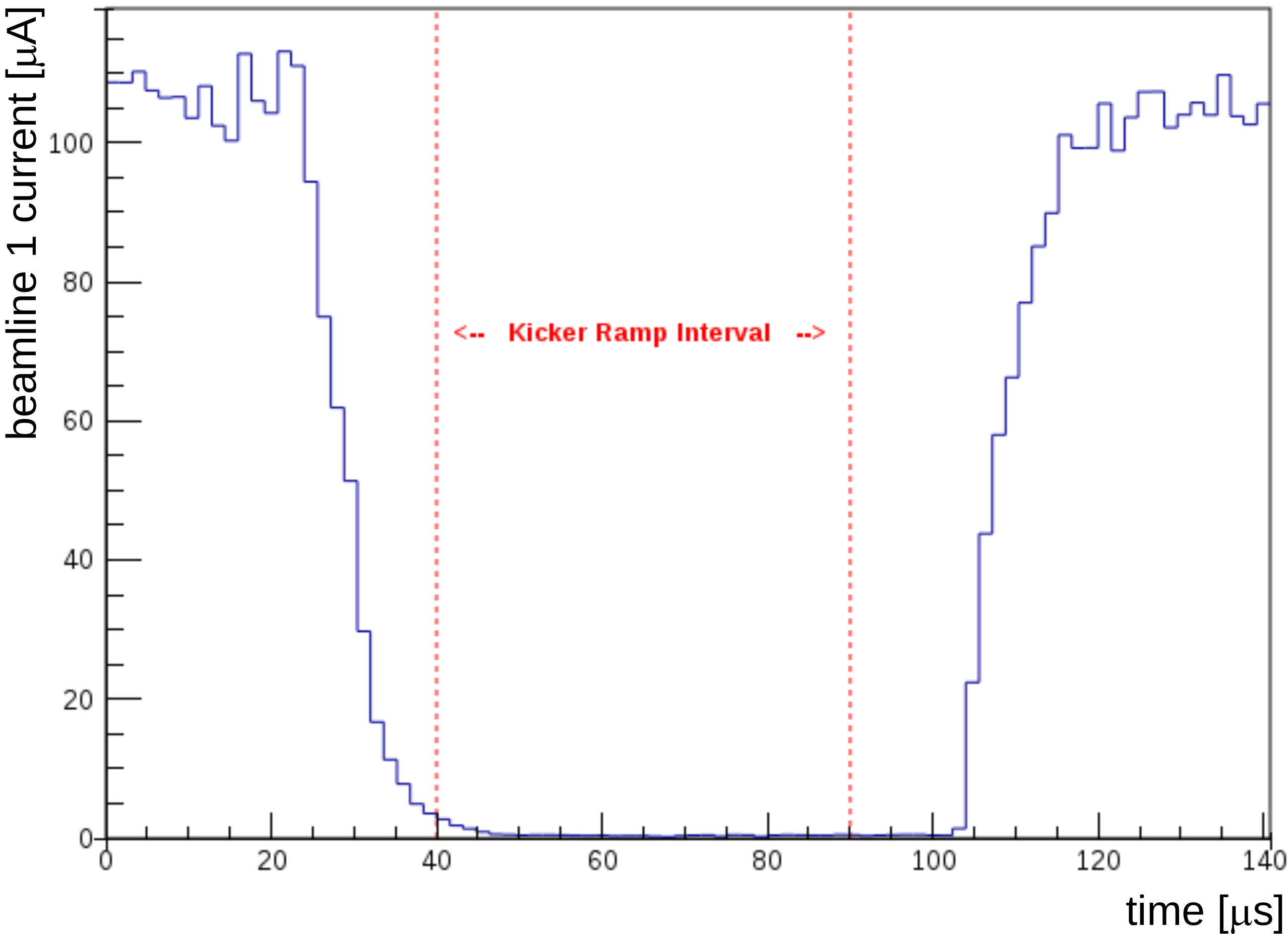} 
	\caption{Typical display of the capacitive probe 1VM4 readout while beamline 1V is at \SI{120}{\micro\ampere}. The beam current is indicated by the continuous line in blue. The beam-off period is long enough for the kicker to ramp up or down during the interval between the vertical, dashed, red lines.}
	\label{fig:notchmonitor}
\end{figure}
To measure the properties and timing of the beam-off period in beamline 1, a capacitive probe 1VM4 was installed downstream of the kicker. It picks up the beam's microstructure consisting of bunches with a length of a few nanoseconds and a repetition rate of 23 MHz.
The pick-off signal is fed into a narrow-band amplifier tuned to the second harmonic at 46 MHz.
The output of the amplifier is a sinusoidal wave; 
its amplitude is proportional to the beam intensity.
The amplifier's filter is designed to provide a fast response  to variations in proton-beam intensity with a transient time of \SI{1}{\micro\second}.
With an electronic noise level equivalent to a beam intensity of \SI{150}{\nano \ampere}, this probe can measure the timing of the macro structure of the cyclotron beam as described above.

This probe can be used to determine the length of the beam-off period between the \SI{1.126}{\kilo \hertz} pulses and to set the right timing for when to ramp the kicker.
An ADC samples the 1VM4 signal with one giga samples per second (1 GSps);
the fast sampling allows full characterization of the fast 46MHz sinusoidal wave.
We calculate an estimate of the beam power from the sinusoidal wave in real time using an FPGA on the digitizer board. 
The kicker sequencing module (KSM) sets the delay of the ramp-up and ramp-down of the current with respect to a digital signal coming from the cyclotron pulser.
It also sets the fraction of pulses that are kicked into beamline 1U to adjust its beam current.
A typical 1VM4 online display is shown in Fig.~\ref{fig:notchmonitor}:
the notch in the beam is about \SI{70}{\micro \second} long and has an exponential tail on the falling edge.
The kicker ramp timing is overlaid for convenience.

\subsection{Lambertson Septum}
\begin{figure}
\centering
	 \includegraphics[width=\columnwidth, trim = {0cm 0cm 1.5cm 4.5cm}, clip]{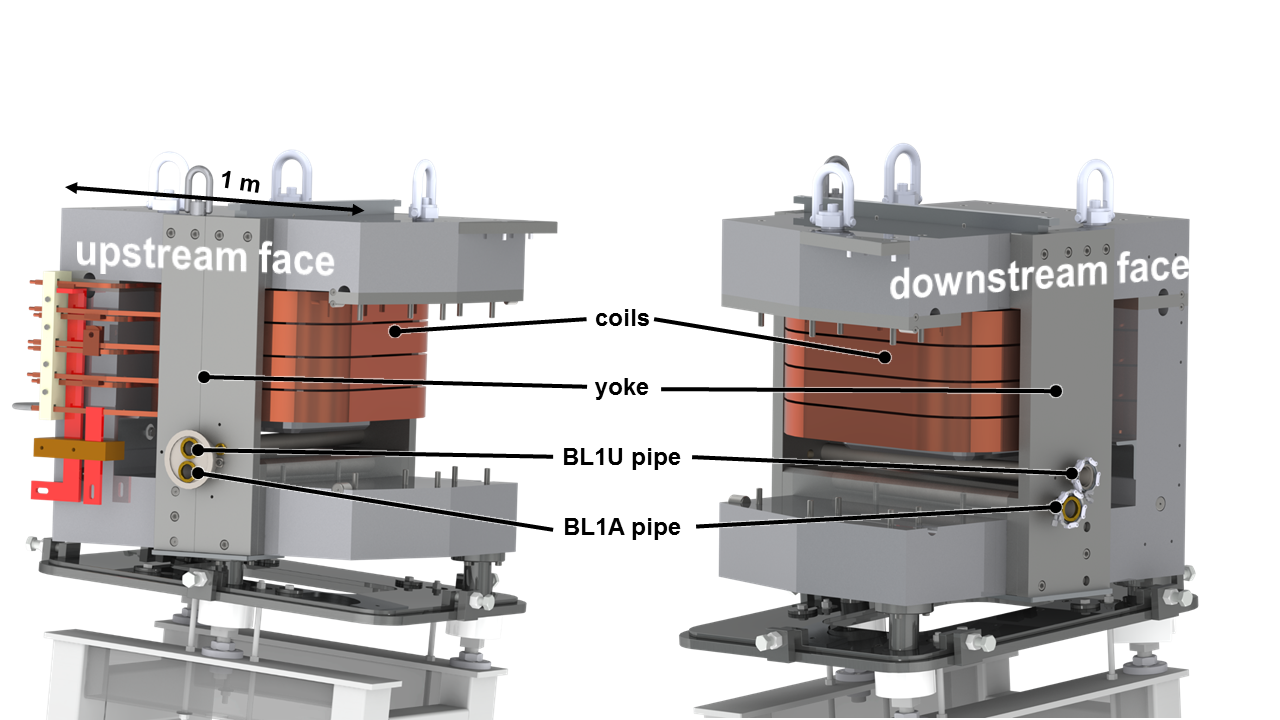} 
	\caption{Rendering of the septum magnet. (left) view of upstream side - (right) view of downstream side. One side of the return yoke has been removed for better visibility of the coils.}
	\label{fig:septum}
\end{figure}
At the location of the Lambertson septum, \SI{5.8}{\meter} downstream of the kicker, the vertical separation between kicked and un-kicked proton beam is \SI{70}{\milli \meter}.
The septum deflects the kicked beam by \SI{150}{\milli \radian} ($9^{\circ}$) to the north while the unkicked beam passes almost undeflected, see Fig.~\ref{fig:BL1Ulayout}.
As shown in Fig.~\ref{fig:septum}, an existing dipole magnet~\cite{ParityDipolePhysRev} has been modified to house two beam tubes.
These have a smaller aperture of \SI{48}{\milli \meter} 
compared to 97~mm for standard proton beamline tubes at TRIUMF.
The upper tube sees the nominal dipole field of \SI{1}{\tesla} (field integral $0.559\,$T$\cdot$m), while the lower is embedded in the pole shoe of the magnet's iron yoke and is virtually field free (field leakage integral $6.9 \times 10^{-4}\,$T$\cdot$m).
This small field leakage causes a deflection of the 1A protons of around \SI{0.2}{\milli \radian} that is easily corrected by the beamline 1A automatic steering elements downstream.

To correct for the upward tilt of the proton beam, the Lambertson septum is rotated around the beamline-1A axis by $4.4^{\circ}$.
At the exit of the septum, the 1U beam is nearly horizontal again, but 80.8~mm higher than the 1A beam.
Any remaining vertical momentum in the 1U beam can be corrected by a small correction bender (1UCBY0) \SI{3.7}{\meter} downstream of the septum magnet.

\subsection{Bending Dipole}
To create more space between the 1U spallation target and  the 1A beamline, another bend is required in beamline 1U.
An additional deflection of \SI{120}{\milli \radian} ($7^{\circ}$) leftward is achieved by the small dipole magnet 1UB0 located 5.62~m downstream of the septum.
It was built by NEC/TOKIN and has a maximum field of around \SI{0.3}{\tesla} at our nominal operating current of around \SI{250}{\ampere}.

\subsection{Quadrupole steering doublet}
\begin{figure}
\centering
	 \includegraphics[width=0.46\columnwidth, trim = {0cm 0cm 0cm 0cm}, clip]{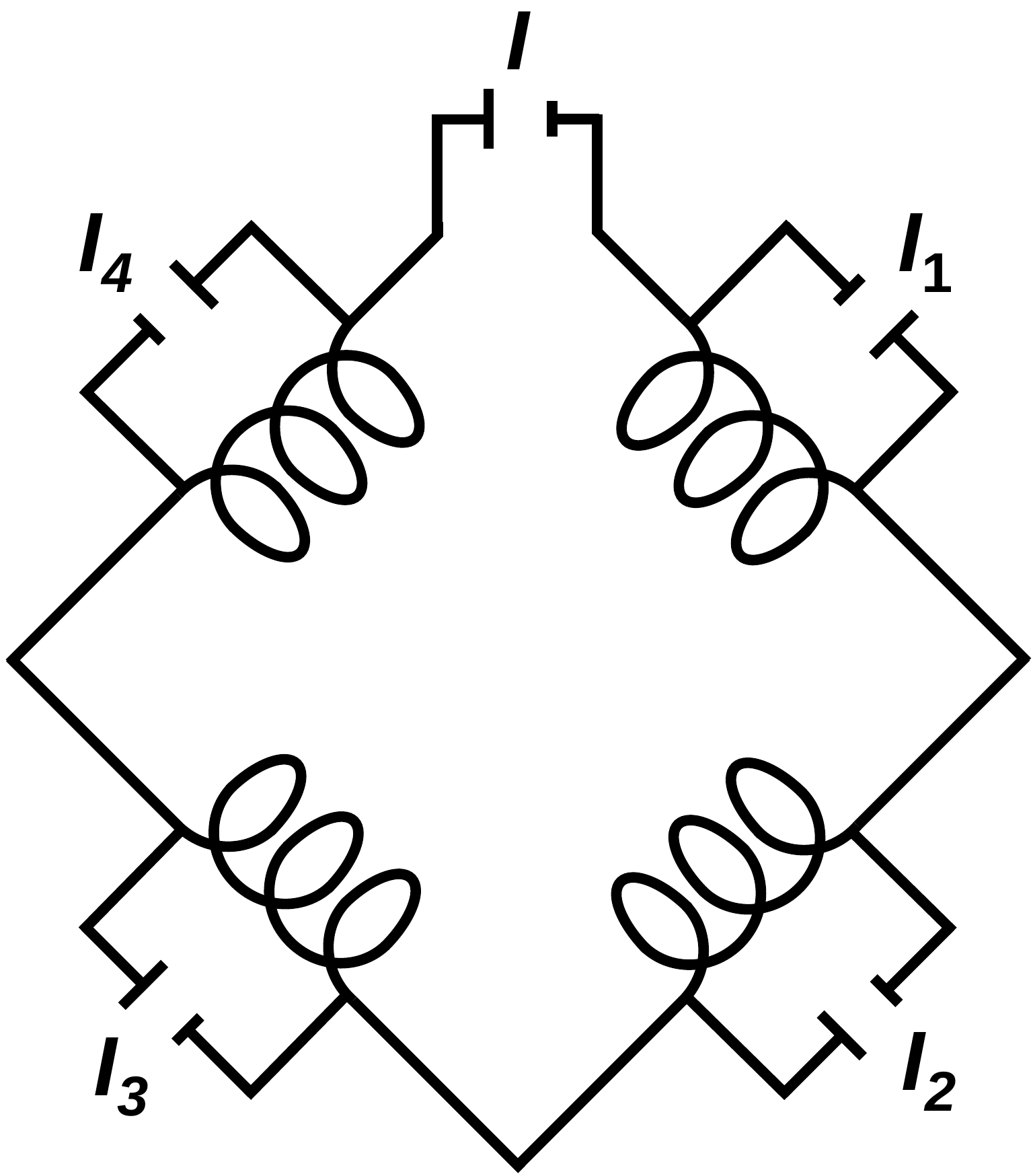} 
     \includegraphics[width=0.34\columnwidth, angle =45,trim = {0cm 0cm 0cm 0cm}, clip]{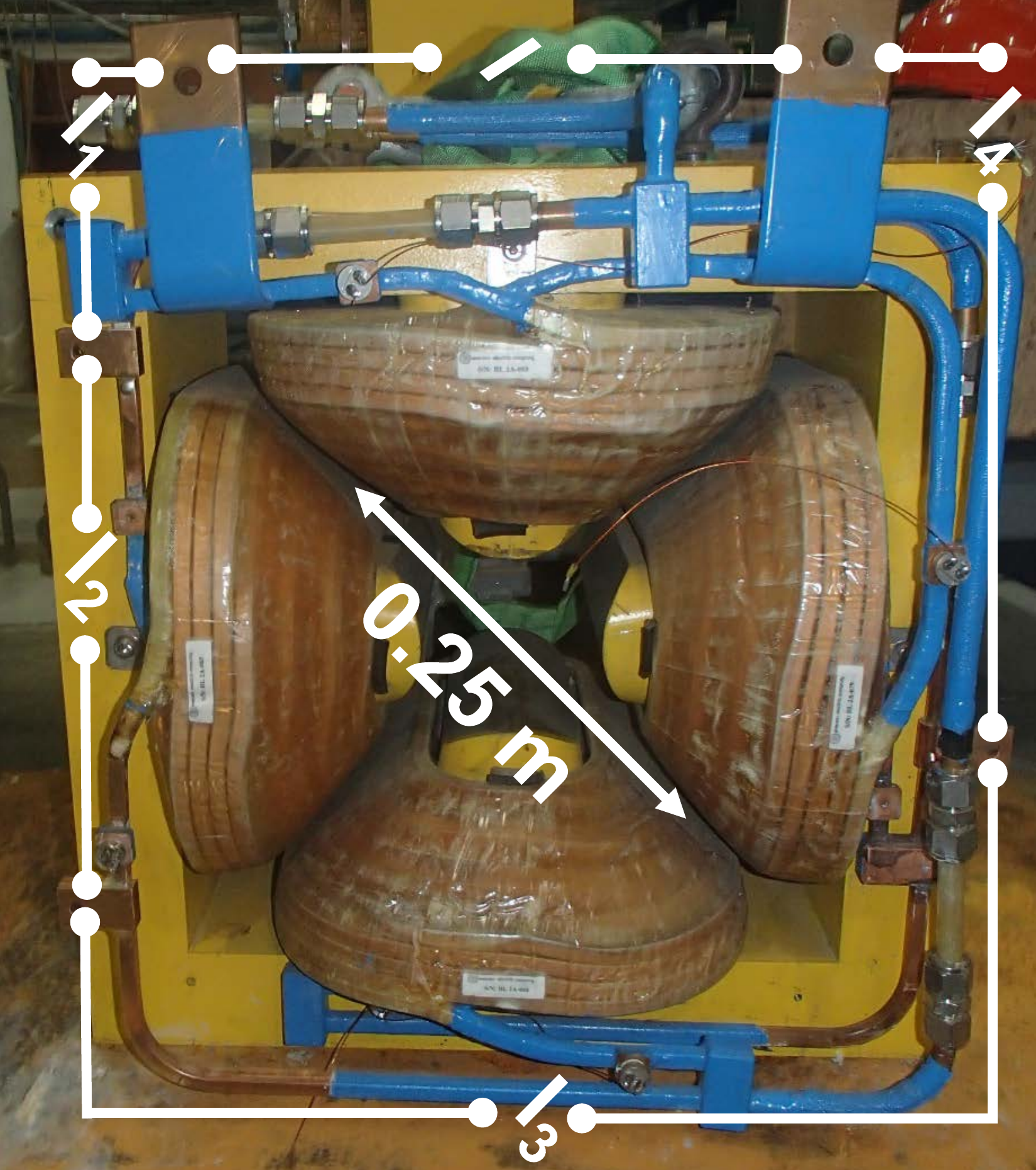} 
	\caption{(left) Powering scheme for the asymmetric steering algorithm for 1UQ1. - (right) Quadrupole magnet 1UQ1 with additional connection taps.}
	\label{fig:1UQ1}
\end{figure}
Two 4Q8.5/8.5-type quadrupole magnets (1UQ1 and 1UQ2) are located closely together downstream of the immovable block.
The quads have a maximum field of \SI{0.9}{\tesla} and the first one is doubling as a steering magnet using an asymmetric steering algorithm.
As shown in Fig.~\ref{fig:1UQ1}, this is achieved by energizing the four coils of the quadrupole with different currents and  superimposed additional dipole field components upon the quadrupole field~\cite{KOLTAY1965,draper1966}.
It effectively shifts the magnetic axis of the quadrupole away from the symmetry point in the center.

Grime~\cite{GRIME2013} describes such a setup using four power supplies.
For technical reasons, we are using a main power supply that energizes all four quadrants in series and four additional smaller power supplies that add current to each individual coil.
In this case, the quadrupole's current can be represented as 
\begin{equation} \label{currentmatrix}
 A = \begin{bmatrix}
    +I + I_{1}       & -I - I_{4} \\
    -I - I_{2}       & +I + I_{3}\\  
\end{bmatrix},
\end{equation}
where $I$ denotes the current from the main power supply and $I_{\rm k} (k=1,2,3,4)$ denotes the small currents from the trim power supplies.
A plus sign in this notation means that the current is flowing in the same direction as for a magnetic North pole.
All $I$ and $I_{\rm k}$ are positive.
Using the superposition principle for magnetic fields, one can decompose the magnet into three components, quadrupole $Q$, horizontal steering $H$ and vertical steering $V$ to give:
\begin{equation} \label{componentmatrix}
 A = \begin{bmatrix}
    +Q       & -Q \\
    -Q       & +Q\\  
\end{bmatrix}
+
\begin{bmatrix}
    -V       & +V \\
    -V       & +V\\  
\end{bmatrix}
+
\begin{bmatrix}
    -H       & -H \\
    +H       & +H\\  
\end{bmatrix}
\end{equation}
Setting expressions~\ref{currentmatrix} and \ref{componentmatrix} equal results in four equations and five degrees of freedom
\begin{align*} 
I_{2} &= 2V+I_{1} \\
I_{3} &= 2H+2V+I_{1} \\
I_{4} &= 2H+I_{1}\\
I &= Q-H-V-I_{1}.
\end{align*}
There are many solutions for these equations, resulting in the desired quadrupole and steering components.

\subsection{Raster magnet}
At the full current of \SI{40}{\micro \ampere}, a small proton beam spot can cause excessive heating of the beamline end window (see section~\ref{sec:window}) and lead to its failure.
Depending on the cyclotron and BL1 beam tune, the $2\sigma$ beam envelope on the window can be circular and as small as \SI{7}{\milli \meter} in diameter. 
A horizontal raster magnet (1URM2) will be installed prior to operation at these high currents.
It will sweep the beam \SI{6.4}{\milli \meter} left and right lowering the time averaged beam intensity.
Two options are currently investigated: either operating the raster magnet at half the kicker frequency or operating it in the low Hz region.

\subsection{Collimator and beamline end window} \label{sec:window}
\begin{figure}
\centering
    \includegraphics[width=\columnwidth, trim = {5cm 0cm 6cm 0cm}, clip]{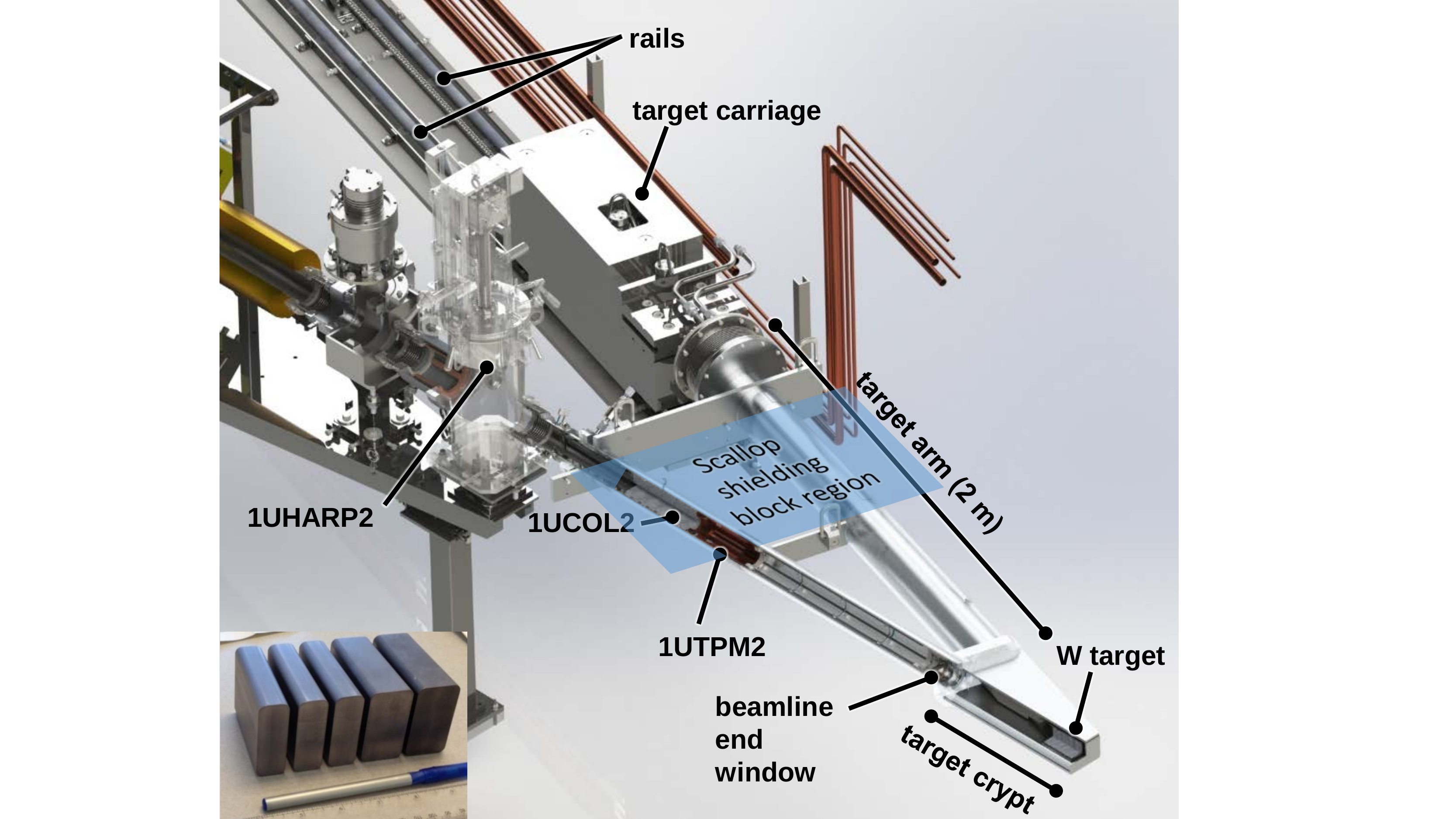} 
	\caption{BL1U end zone, spallation target assembly and target extraction mechanism. Inset: tungsten target blocks.}
	\label{fig:target}
\end{figure}
\begin{figure}
\centering
     \includegraphics[width=\columnwidth,  trim = {5cm 1cm 6cm 4.2cm}, clip]{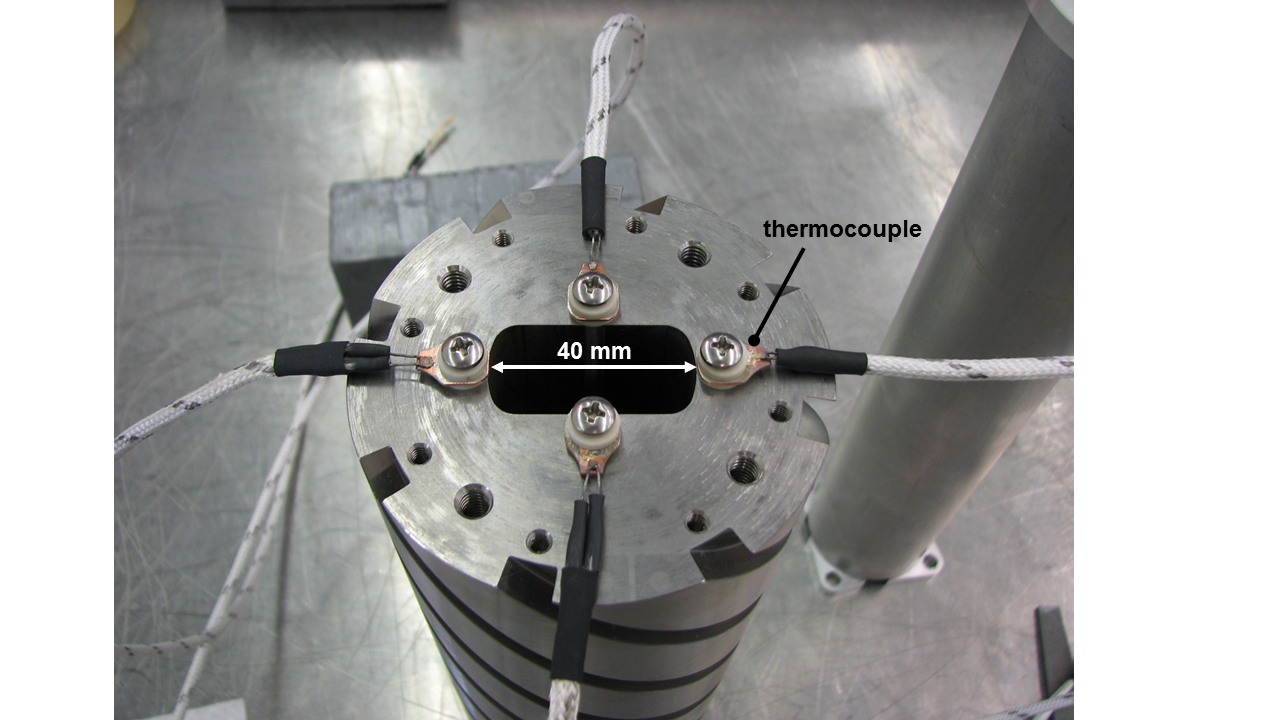} 
	\caption{Photo of tungsten collimator, showing its upstream face including the four temperature sensors.}
	\label{fig:collimator}
\end{figure}

The in-pile section of BL1U (see Fig.~\ref{fig:target}) houses a 30~cm-long tungsten collimator, 1UCOL2, with a 20 x \SI{40}{\milli \meter} opening.
The front face of the collimator is equipped with four thermocouple sensors, one in each quadrant of the collimator (up, left, down, right) see Fig.~\ref{fig:collimator}.
These thermocouples provide a good handle on the beam position since they provide a larger signal when hit by beam halo.

\begin{figure}
\centering
    \includegraphics[width=\columnwidth, trim = {0cm 4.9cm 0cm 1cm}, clip]{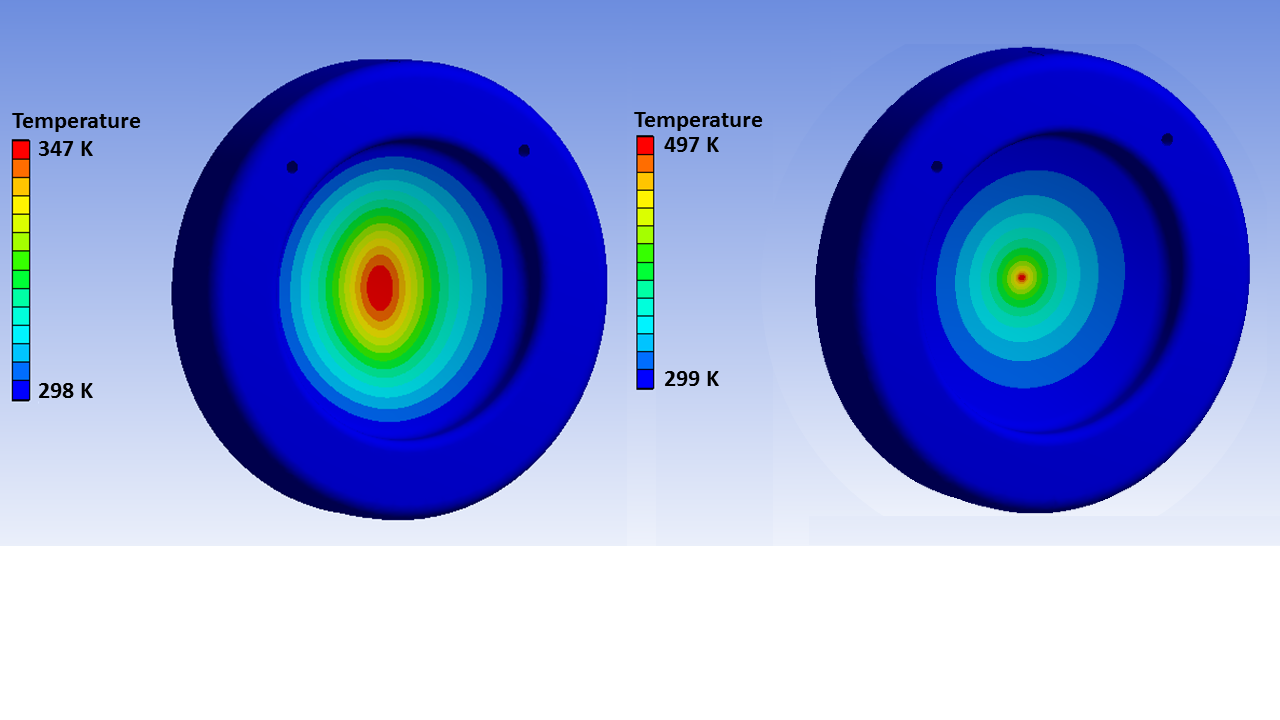} 
	\caption{Left: temperature profile in Kelvin calculated with ANSYS for steady operation of the beamline end window at \SI{40}{\micro\ampere} with a Gaussian, elliptic beam spot of $34 \times14$~mm$^2$ at 2$\sigma$. - Right: temperature profile for steady operation of the beamline end window at \SI{40}{\micro\ampere} with a Gaussian beam spot of $1 \times 1$~mm$^2$ at 2$\sigma$.}
	\label{fig:windowtemps}
\end{figure}

At the end of BL1U, a \SI{1}{\milli \meter}-thick window machined out of a full aluminum block separates the vacuum space from the target crypt.
The crypt is filled with helium at a pressure of approximately \SI{350}{\milli \bar} above atmosphere.
Aluminum 2219 T851 has been chosen due to its strength at elevated temperature and its reasonable thermal conductivity.
Heat deposit to this window is roughly \SI{25}{\watt} at nominal beam current.
Convection in the helium on the downstream side and a water-cooling jacket at the flange surrounding the window provide cooling.
Fig.~\ref{fig:windowtemps} shows that a larger Gaussian beam spot of $34 \times14$~mm$^2$ at 2$\sigma$ can be sustained, while the worst case scenario beam spot ($1 \times 1$~mm$^2$ at 2$\sigma$) heats the window far above the creep temperature limit of around 100$^\circ$C for aluminum 2219 T851.
If this small beam spot is not rastered across the target, window failure will result.

\subsection{Spallation target}
We chose a tungsten target for the spallation source at TRIUMF:
MCNP calculations show that the fraction of neutron yield divided by heat load to the UCN production volume is comparable to pure tantalum or lead.
Cooling water corrodes tungsten, therefore we followed a design based on KENS, the spallation neutron source at High Energy Accelerator Research Organization (KEK)~\cite{KENS} that has proven to be reliable: the blocks were clad in tantalum with a thickness \SI{> 0.1}{\milli \meter} using hot, isostatic pressing (HIP), a high-pressure and high-temperature process in which the two metals are fused~\cite{KENS,LANSCETaW}. 

As shown in Fig.~\ref{fig:target}, the target is  made up of five blocks of tungsten;
they are \SI{78}{\milli \meter} high and \SI{57}{\milli \meter} mm wide, three are \SI{20}{\milli \meter} long in beam direction and two are \SI{30}{\milli \meter} long.
A water flow of approximately 0.8 liter per second cools the target.
Horizontal channels around the blocks create a uniform flow.
\begin{figure}
\centering
	 \includegraphics[width=\columnwidth, trim = {0cm 1cm 1cm 4cm}, clip]{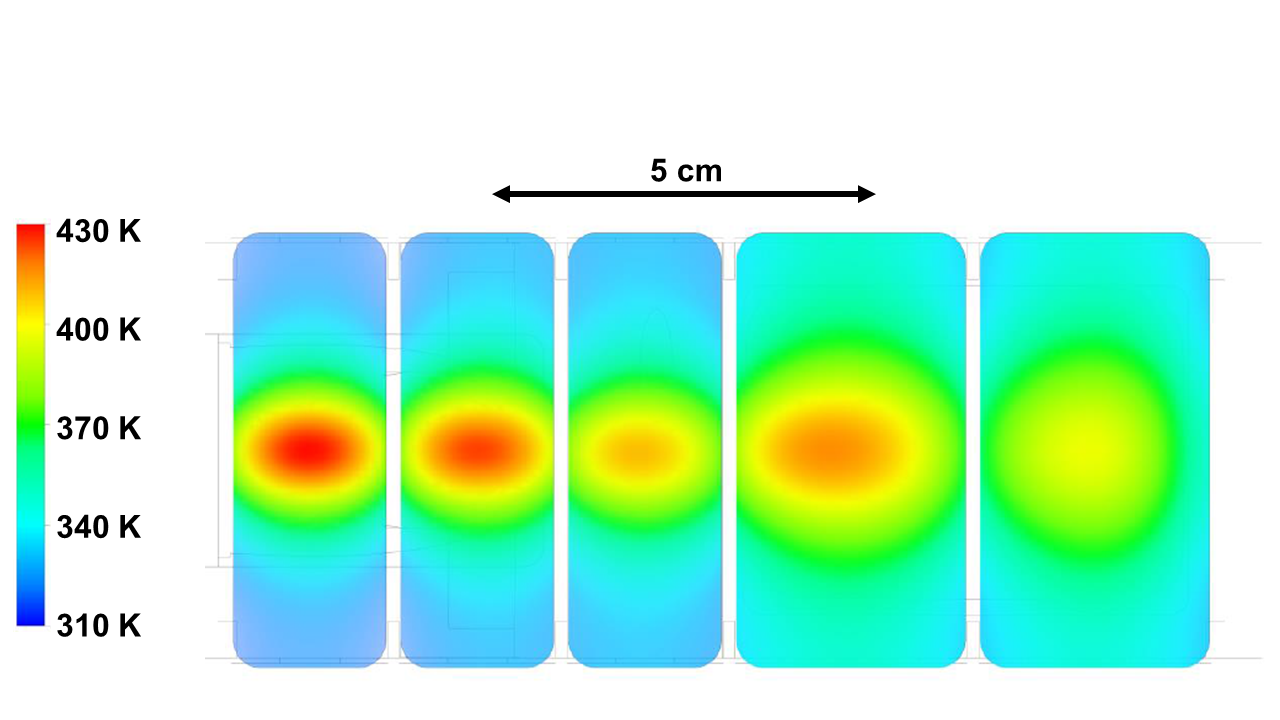} 
	\caption{Temperature distribution at the vertical center of the spallation target blocks as calculated by an ANSYS~\cite{ansys} simulation.
An elliptical beam profile of \SI{10}{\milli \meter} height and \SI{20}{\milli \meter} width ($2\sigma$) at \SI{520}{\mega \electronvolt} and a proton current of \SI{40}{\micro\ampere} were used in an MCNPX study to generate heat input profiles for ANSYS CFX, where a thermal analysis was performed.}
	\label{fig:targettemps}
\end{figure}
The temperature profile during irradiation can be seen in Fig.~\ref{fig:targettemps}.
It visualizes the rationale for using two different block thicknesses.
Most of the heat is deposited in the upstream part of the target, therefore more cooling is necessary in this region.
This is achieved by making the first three blocks thinner generating more cooling channels along the beam direction and therefore less stress in this region.
To reduce beam absorption in water which has lower neutron yield than tungsten, the last two blocks are thicker.

The estimated lifetime of the target is longer than ten years at \SI{40}{\micro \ampere} beam operation during approximately eight months of the year.
At its end of life, the target has to be replaced.
To this end, the target is mounted on a target arm that is cantilevered from a carriage running on rails, see Fig.~\ref{fig:target}.
The carriage can be retracted after the biological shielding has been removed and service connections to the target were disconnected.

We performed MCNP simulations irradiating the target with 480-MeV protons for ten years with an average current $\bar{I}=$\SI{6.7}{\micro \ampere}.
The average current is estimated by
\begin{equation}
\bar{I}= I_{\rm BL1} \times T_{\rm up} \times f \times d,
\end{equation}
where $I_{\rm BL1} =$\SI{120}{\micro \ampere} denotes the average proton current in beamline 1, $T_{\rm up} = 8/12$ the maximum uptime fraction of the cyclotron taking the annual cyclotron shutdown into account, $f = 1/3$ the kick fraction and $d = 0.25$ the duty cycle for UCN experiments;
a usual UCN experiment will need around one minute of beam on to generate enough UCN, then store the neutrons for several minutes and then count them (we assume three minutes).
The dose rate at a distance of \SI{50}{\centi \meter} distance from the target after ten years of operation is \SI{4.4}{\sievert \per \hour} after a cooldown period of one hour and \SI{1.8}{\sievert \per \hour} after one week.
To transport and store the target, a fifteen centimeter thick lead casket will enclose the target.
This will reduce the dose rate \SI{50}{\centi \meter} distance to \SI{93}{\micro \sievert \per \hour} after one week of cooldown and makes it manageable for the remote handling group.

\subsection{Diagnostic elements}
Three wire chambers~\cite{HARPS} are installed as beam profile monitors: 1AM5 before the septum, 1UHARP0 before the bender 1UB0 and 1UHARP2 before the collimator.
The HARPS are gas-filled and contain wire planes for the vertical and horizontal profiles of the beam. Wire separation is 3~mm.
They can only be used during low current operation of less than 50~nA due to excessive beam spills caused by proton scattering.

The beam current in BL1U is measured by a toroidal, non-intercepting, transformer-style beam monitor, 1UTNIM2.
It provides a beam-interlock signal in case of over-current to protect the beamline window and target~\cite{beamdiag}.
This is a licensing requirement from the Canadian Nuclear Safety Commission.
Special electronics, several filter stages and PLC averaging are used to ensure that the TNIM signal is linear and provides a good beam estimate from \SI{0.1}{\micro \ampere} to \SI{40}{\micro \ampere}.

Two beam spill monitors (BSM55 and BSM56) are installed along the beamline, one downstream of the septum and one downstream of the two quadrupole magnets.
In case of excessive beam spill, they provide an interlock to shut down 
the cyclotron.

Two inductive, non-intercepting beam position monitors (1UBPM2.1 and 1UBPM2.2) are installed downstream of the last quadrupole~\cite{triumfbpm2007}.
They will measure the horizontal and vertical center of gravity of the proton beam.

The last diagnostic element right before the beamline end window is a secondary-electron-emission-type target protect monitor (1UTPM)~\cite{TPM}.
It features split tantalum foils of \SI{5}{\milli \meter} active width around a 40~mm by 15~mm aperture to measure the beam halo in four quadrants.

\section{Beam optics} \label{opticscalc}
\begin{figure*}
\centering
	 \includegraphics[width=\textwidth]{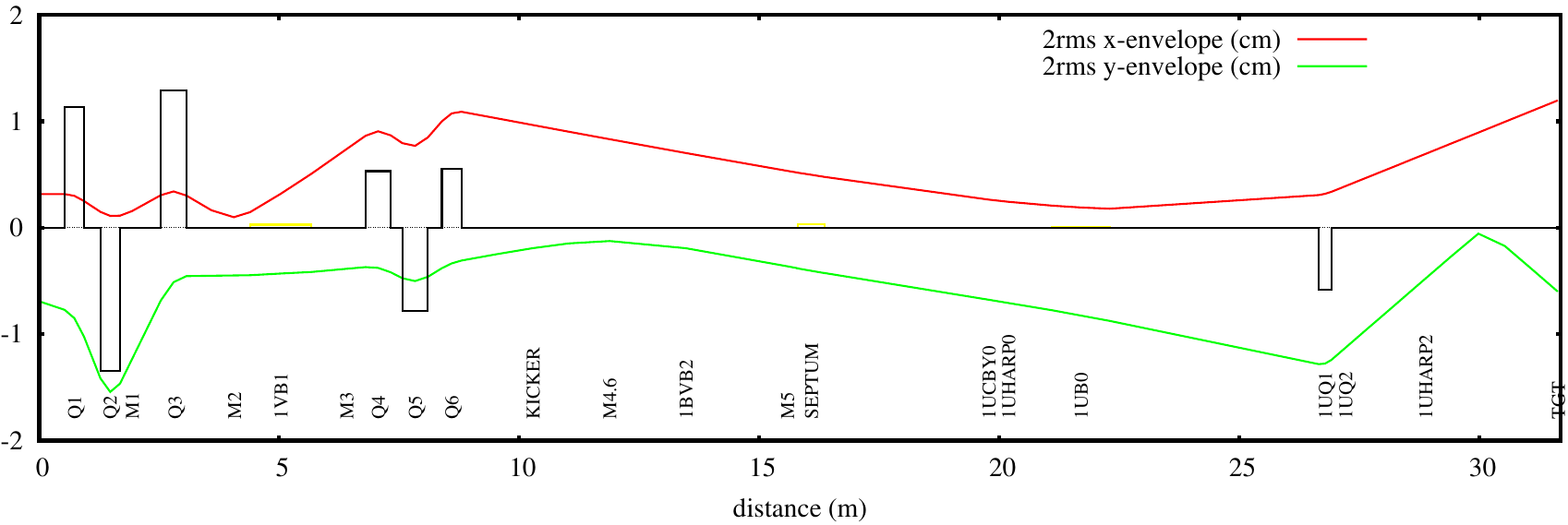} 
	\caption{Beam envelope plot of beamline 1V and 1U calculated with Transoptr\cite{transoptr}. The black bars denote the magnetic field created by the optics elements.}
	\label{fig:envelope}
\end{figure*}
As shown in Fig.~\ref{fig:BL1Ulayout}, beamline 1U shares some beam optics with the vault section of beamline 1 (called 1V) up to the septum magnet.
The main task of the front-end triplets 1VQ1 to 1VQ6 is to provide a small beam size (\SI{<1}{\centi \meter} at 2~$\sigma$) and small dispersive elements $R_{16}$, $R_{26}$ in the beam transport matrix downstream up to the septum.
A small beam size is necessary to reduce spill on the smaller beam tubes of the septum.
Additionally, the triplets allow to adjust the beam spot on the BL1U target in horizontal and vertical direction with little crosstalk.
During normal cyclotron operation, the momentum spread of the beam coming out of the cyclotron is smaller than 0.5\% (2~$\sigma$). 
Fig.~\ref{fig:envelope} shows a beam envelope calculation of beamlines 1V and 1U.
As discussed above, the beam spot size on the beamline end window and target is critical and a raster magnet is required if the beam spot is too small.
1UQ1, 1UQ2 and 1VQ6 can be tuned to create a horizontally wide spot at the UCN target, because at 1VQ6 the beam horizontal
size is significantly larger than the vertical one, making the horizontal size at the UCN target sensitive to any change in 1VQ6.
Therefore, the horizontal beam profile might become asymmetric because both the momentum spread and the initial image of the beam at the stripping foil are asymmetric.

\section{Neutron moderators and UCN source} \label{source}
\begin{figure}
	\centering
	 \includegraphics[width=0.46\columnwidth, trim = {4cm 0cm 1cm 0cm}, clip]{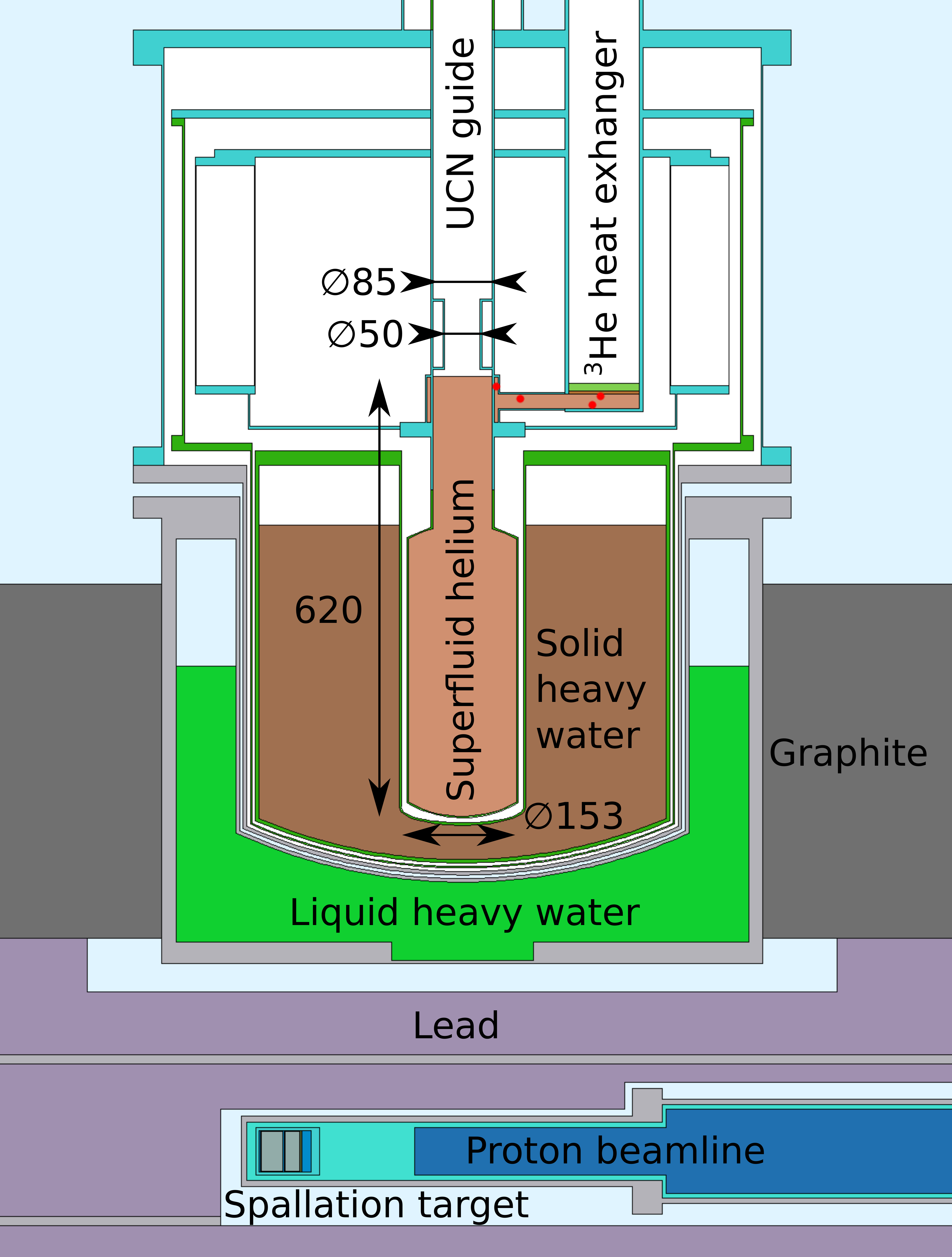} 
	 \includegraphics[width=0.52\columnwidth, trim = {6.5cm 0cm 13cm 0cm}, clip]{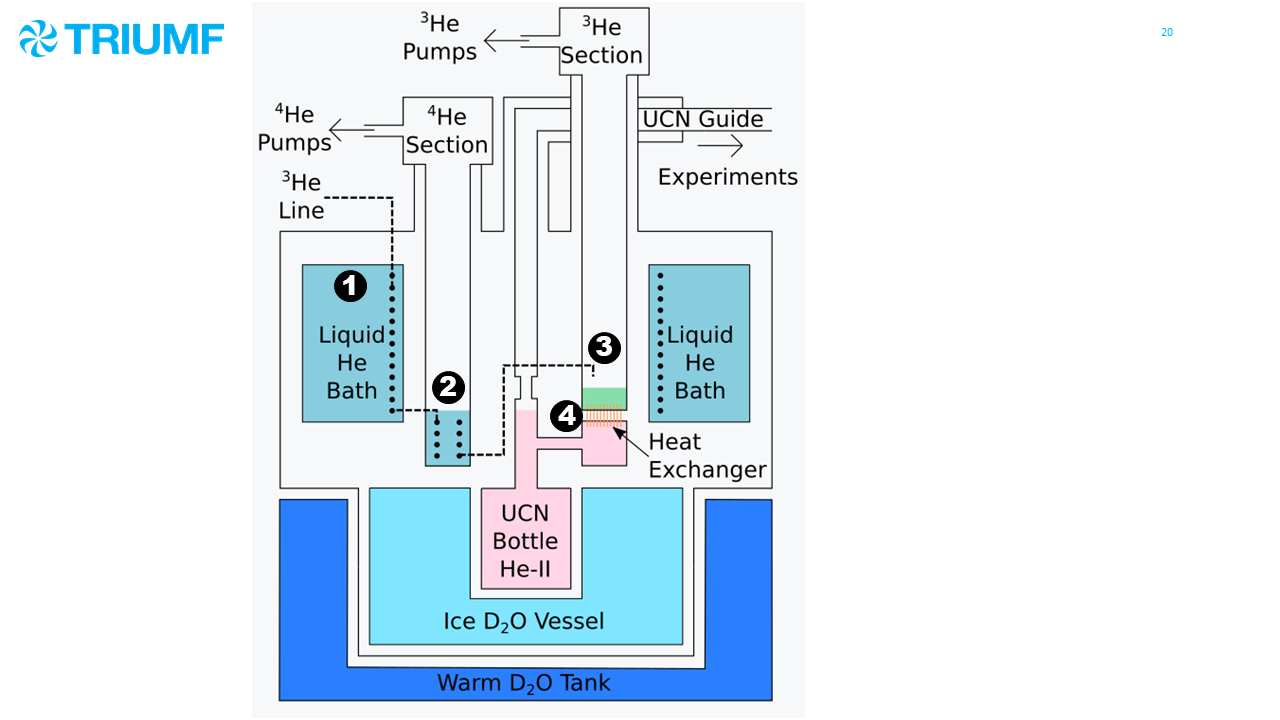} 
	\caption{(left) Cross section of the moderator and UCN source installed above the spallation target. - (right) The four stages of the UCN source cryostat are indicated by the numbers, see text.}
	\label{fig:source}
\end{figure}
In 2017, we installed a prototype UCN source above the spallation target.
This source has been developed in Japan and extensively tested at RCNP, Osaka~\cite{rcnp} and has a cooling power compatible with a proton beam operation of several \SI{}{\micro\ampere}.
Fig.~\ref{fig:source} illustrates how the UCN production scheme mentioned in the introduction is currently implemented at TRIUMF:
the spallation target crypt is surrounded by lead to shield the cryostat from gamma rays.
Liquid heavy water at room temperature and solid heavy water at around \SI{15}{\kelvin} serve as neutron moderators.
Graphite surrounding the source reflects a portion of the escaping neutrons back to the moderator and converter region.

A cryocooler is used to cool down the heavy water.
The superfluid helium converter is cooled down to  \SI{< 1}{\kelvin} using a four-stage cryostat.
We will describe the main components here:
natural liquid helium from a liquefier facility in the Meson hall at TRIUMF supplies the  liquid helium bath (stage 1).
A part of this liquid is filled into a natural helium pot that is being pumped on to lower its temperature to about \SI{1.5}{\kelvin} (stage 2).
Heat exchanging coils in this reservoir pre-cool and condense $^3$He which is filled into a $^3$He pot where large pumps lower the pressure to reach temperatures as low as \SI{0.72}{\kelvin} (stage 3).
The superfluid helium in the production volume has to be isotopically pure ($^3$He impurity levels lower than $10^{-10}$) to avoid excessive UCN losses via absorption.
Hence, it has to be cooled down via a $^3$He-$^4$He heat exchanger (stage 4).
The cooling power of the cryostat is around \SI{300}{\milli \watt} at \SI{0.9}{\kelvin}.
UCN created in the production volume are extracted to experiments via material guides of adequately high Fermi potential (aluminum coated with nickel-phosphorus and stainless steel).

The prototype source does not have enough cooling power to keep the isopure helium cold enough for efficient UCN production at the nominal beamline 1U current of \SI{40}{\micro\ampere}.
Since the cryostat is located right above the spallation target, radiation damage of its components would also not allow long term operation at this current.
An upgrade of the UCN source is underway to exploit the full potential of the beamline and spallation target.
It will feature a liquid deuterium moderator to increase the cold neutron flux compared to solid heavy water.
Both, the liquid deuterium and helium cryostats will be located far enough away from the spallation target to allow long term operation.
We are designing the helium cryostat for an increased cooling power of around \SI{10}{\watt} at a slightly higher operating temperature of \SI{1.1}{\kelvin}.
Commissioning of this new source is planned for 2021.

\section{Commissioning with protons} \label{commissioning}

\paragraph{Commissioning with continuous beam}
\begin{figure}
	\centering
	 \includegraphics[width=\columnwidth]{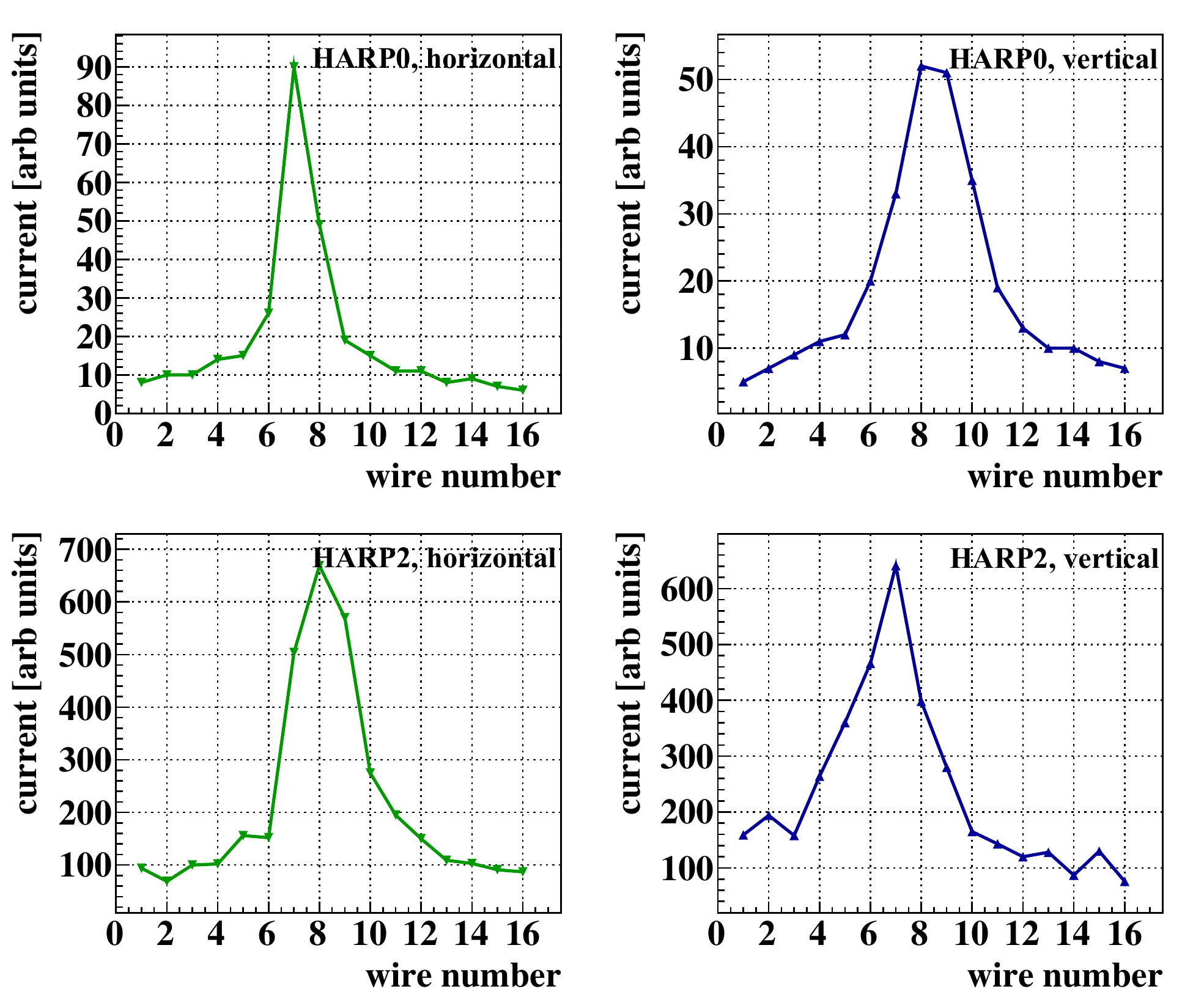} 
	\caption{Well centered beam profiles on B1U:HARP0 (top) and HARP2 (bottom) during BL1U commissioning in November 2016. Left: horizontal profile, right: vertical profile.}
	\label{fig:HARPSNov2017}
\end{figure}
The beamline 1U hardware was installed during the TRIUMF main cyclotron shutdown periods 2014, 2015 and 2016.
First beam was injected into BL1U on November 22, 2016. 
Within one hour, beam tuning could be completed.
Fig.~\ref{fig:HARPSNov2017} shows the beam profile at the two HARPs during this day at less than \SI{1}{\nano \ampere} of proton current. 
Beam on target was confirmed by a temperature increase of \SI{4}{\kelvin} in the target cooling water caused by a \SI{1}{\micro\ampere} proton beam.
During these first beam periods, the kicker magnet was set to nominal current before beam was injected into 1V and was being kept energized in DC mode throughout the tests so that the whole beam in 1V was diverted to 1U.

\begin{figure}
	\centering
	 \includegraphics[width=\columnwidth, trim = {0cm 0cm 0cm 0cm}, clip]{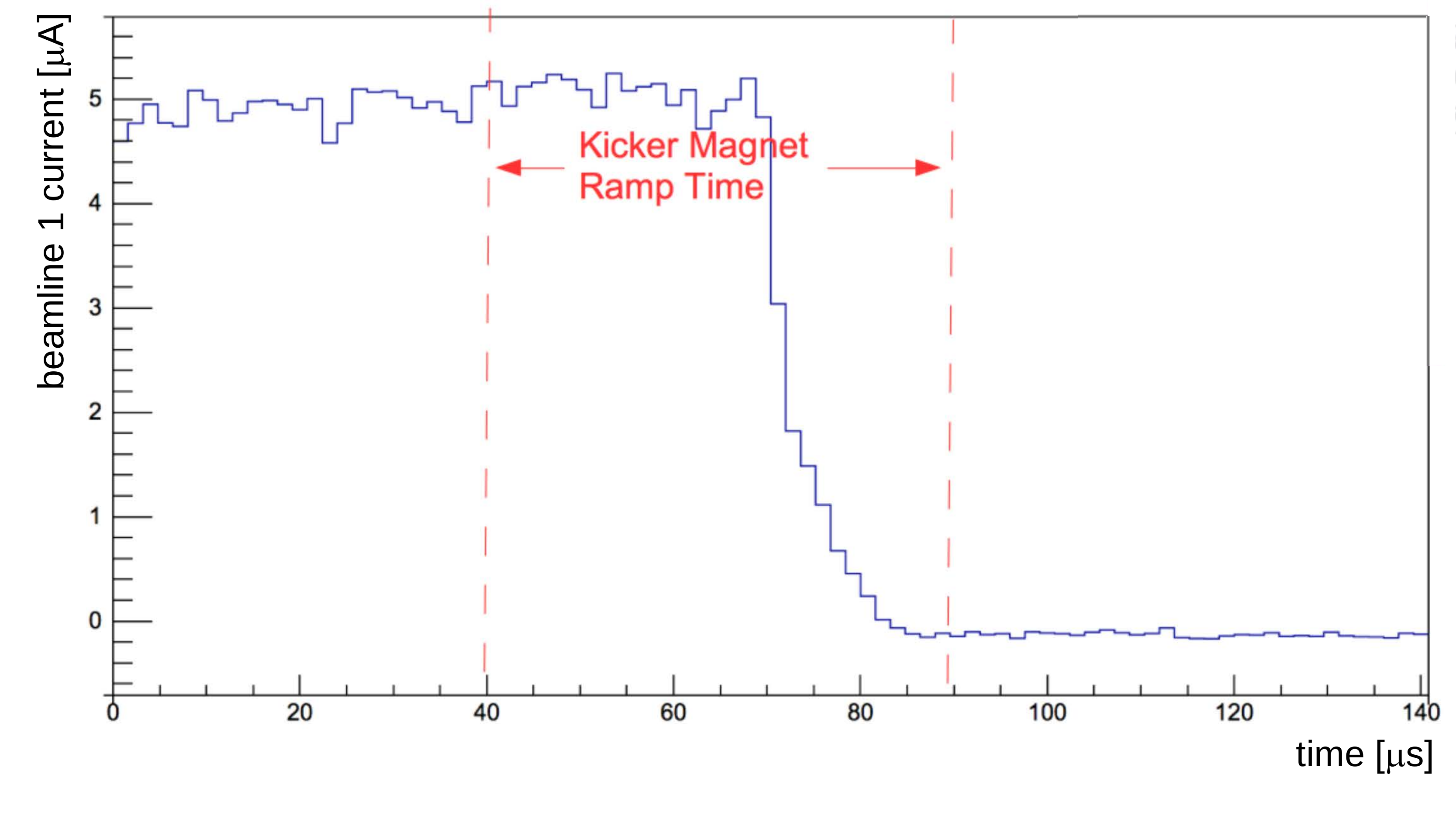} 
	\caption{Overlap of kicker ramp with a beam-on period in beamline 1 at a delay time of \SI{300}{\micro \second}.}
	\label{fig:1VM4mistiming}
\end{figure}

\paragraph{Commissioning with shared beam}
\begin{figure}
	\centering
	 \includegraphics[width=0.65\columnwidth, trim = {0cm 0cm 0cm 0cm}, clip]{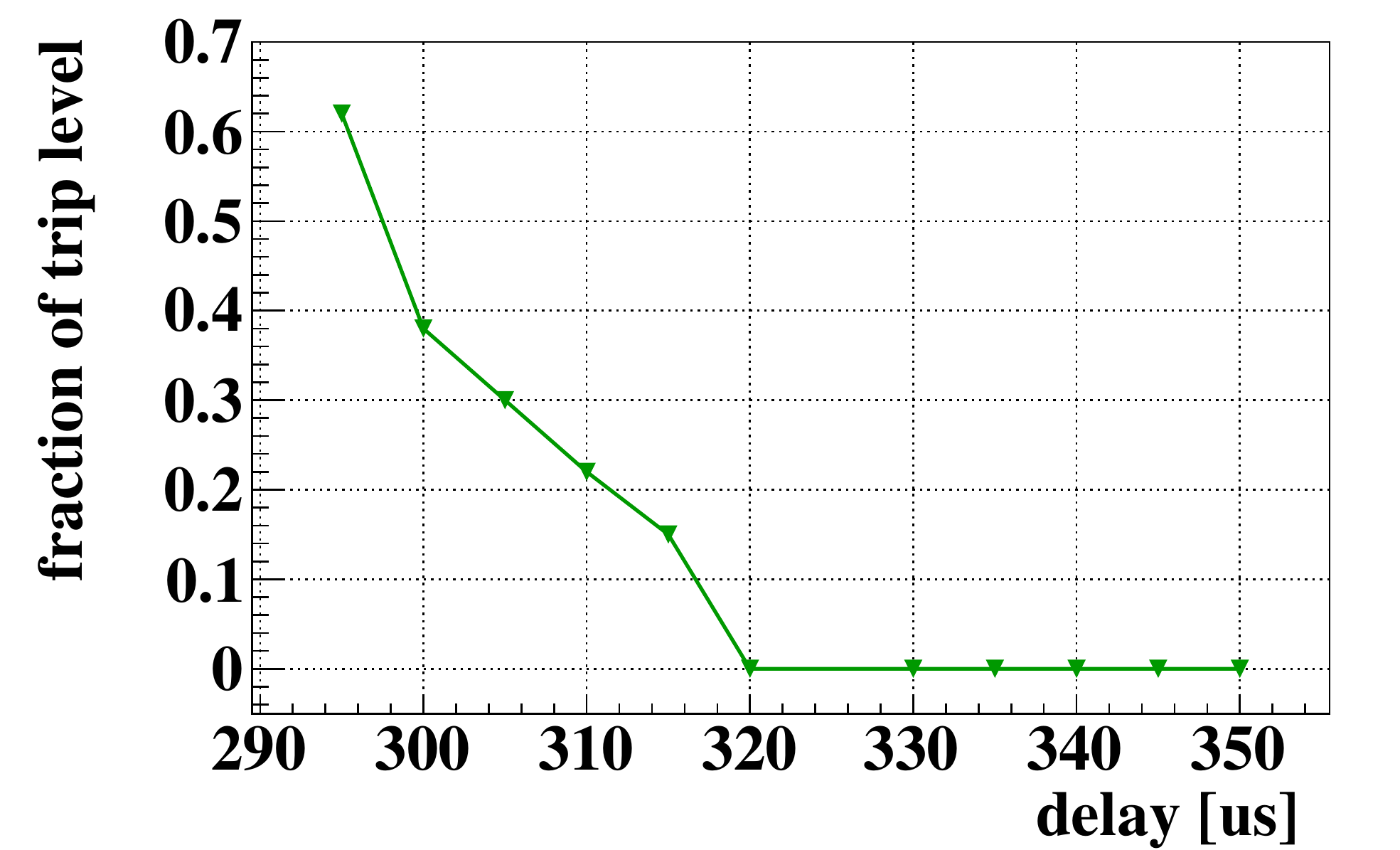} 
	\caption{Signal of the beam spill monitor 55 located downstream of the septum yoke for different kicker delays with respect to the cyclotron pulser signal.}
	\label{fig:kicktiming}
\end{figure}
The capacitive probe 1VM4 was commissioned in May 2017.
With the capacitive probe and kicker sequencing module operational, it became possible to share beam between beamlines 1A and 1U.
In June 2017, first kicker tests were performed at low current of \SI{1}{\micro\ampere} in 1V.
The beam spill monitor 55 downstream of the septum magnet was used to understand the consequences of mis-timing the kicker ramp.
Fig.~\ref{fig:kicktiming} shows the beam spill as a function of the delay of the kicker ramp to the pulser signal.
Most of the delay comes from the transit time of protons.
Lowering the kicker delay relative to the cyclotron pulser signal below \SI{320}{\micro \second} causes significant beam spill on the yoke of the septum magnet, since the kicker is sweeping the beam across this yoke during a time period with significant proton beam.
The capacitive probe shows a large overlap of the kicker ramp interval with a beam-on period at delay of \SI{300}{\micro \second}, see Fig.~\ref{fig:1VM4mistiming}.

In July 2017, beamline 1U began operating with a proton current of \SI{1}{\micro\ampere} by kicking every 120th pulse from around \SI{120}{\micro\ampere} in beamline 1.

\paragraph{UCN Commissioning}
First UCN production with the prototype source was achieved in fall 2017.
The UCN yield and storage lifetime in the source were consistent with expectations from Monte-Carlo simulations:
up to $4.8 \times 10^{4}$ neutrons were detected after irradiating the target for \SI{60}{\second} at a proton current of \SI{1}{\micro\ampere}.
The peak temperature of the superfluid helium in the production volume was around \SI{0.91\pm0.07}{\kelvin}.
Storage lifetimes of up to \SI{ 37.30\pm0.09}{\second} were observed.
During test runs, the current was also ramped up to \SI{10}{\micro\ampere}:
no issues with the beamline end window or target cooling were discovered.
The UCN yield increased up to $3.3 \times 10^{5}$ UCN for a \SI{60}{\second} at this current while the peak temperature of the superfluid helium in the production volume increased to \SI{1.12}{\kelvin}.
More details can be found in~\cite{UCNprod}.

In 2018, the number of UCN detected for a \SI{60}{\second}, \SI{1}{\micro\ampere} irradiation could be increased to $7.2 \times 10^{4}$ due to better coupling of the detector to the UCN source.

\section{Conclusion and outlook}
During the years 2013 to 2016, a new proton beamline and neutron spallation target were installed at TRIUMF.
It features a fast kicker magnet to share beam between the existing beamline 1A with its Center for Molecular and Material Science and the new neutron spallation target.
After successful commissioning in 2016 and 2017, the first cold and ultracold-neutron-production experiments were performed~\cite{UCNprod}.

A raster magnet will be designed and built to allow full current operation of \SI{40}{\micro\ampere} by 2021. 
At this current, beamline 1U at TRIUMF will provide a large enough neutron flux to create a world-leading UCN source to be used for fundamental physics with neutrons.

\section*{Acknowledgements}
We would like to thank C.A.~Miller and C.~Marshall for their invaluable contributions to beamline 1U.

This work was supported by the Canada Foundation for
Innovation, the Natural Sciences and Engineering Research Council of Canada, Research Manitoba and the Japan Society for the Promotion of Science.
The research was undertaken, in part, thanks to funding from the Canada Research Chairs program.

\section*{References}

\bibliography{mybibfile}

\end{document}